%% file: main.tex
\documentclass[preprint,authoryear]{elsarticle}

\usepackage[colorlinks]{hyperref}
\usepackage{graphicx}
\usepackage{amsmath}
\usepackage{hhline}
\usepackage{booktabs}   
\usepackage{multirow}
\usepackage{xcolor}
\usepackage{xurl}
\usepackage{subcaption}
\usepackage{float}

\newcommand{\answer}[1]{#1}

\journal{Medical Image Analysis}

\biboptions{authoryear}

\begin{document}

\begin{frontmatter}

\title{CT-based COVID-19 Triage: Deep Multitask Learning Improves Joint Identification and Severity Quantification}


\author[skoltech,iitp]{Mikhail Goncharov\fnref{equal}}
\author[skoltech,iitp]{Maxim Pisov\fnref{equal}}
\author[iitp]{Alexey Shevtsov\fnref{equal}}
\author[skoltech,iitp]{Boris Shirokikh\fnref{equal}}
\fntext[equal]{Equal contribution}
\author[iitp]{Anvar Kurmukov}
\author[radiology]{Ivan Blokhin}
\author[radiology]{Valeria Chernina}
\author[sklif]{Alexander Solovev}
\author[radiology]{Victor Gombolevskiy}
\author[radiology]{Sergey Morozov}
\author[skoltech]{Mikhail Belyaev\corref{mycorrespondingauthor}}
\cortext[mycorrespondingauthor]{Corresponding author}
\ead{m.belyaev@skoltech.ru}

\address[skoltech]{Skolkovo Institute of Science and Technology, Moscow, Russia}
\address[iitp]{Kharkevich Institute for Information Transmission Problems, Moscow, Russia}
\address[radiology]{Research and Practical Clinical Center for Diagnostics and Telemedicine Technologies of the Moscow Health Care Department}
\address[sklif]{Sklifosovsky Clinical and Research Institute for Emergency Medicine, Moscow, Russia}

\begin{abstract}
The current COVID-19 pandemic overloads healthcare systems, including radiology departments. Though several deep learning approaches were developed to assist in CT analysis, nobody considered study triage directly as a computer science problem. We describe two basic setups: \textit{Identification} of COVID-19 to prioritize studies of potentially infected patients to isolate them as early as possible; \textit{Severity quantification} to highlight studies of severe patients and direct them to a hospital or provide emergency medical care. We formalize these tasks as binary classification and estimation of affected lung percentage. Though similar problems were well-studied separately, we show that existing methods provide reasonable quality only for one of these setups. We employ a multitask approach to consolidate both triage approaches and propose a convolutional neural network to combine all available labels within a single model.  \answer{In contrast with the most popular multitask approaches, we add classification layers to the most spatially detailed upper part of U-Net instead of the bottom, less detailed latent representation.} We train our model on approximately 2000 publicly available CT studies and test it with a carefully designed set consisting of \answer{32 COVID-19 studies, 30 cases with bacterial pneumonia, 31 healthy patients, and 30 patients} with other lung pathologies to emulate a typical patient flow in an out-patient hospital. \answer{The proposed multitask model outperforms the latent-based one and achieves ROC AUC scores ranging from $0.87\pm0.01$ (bacterial pneumonia) to $0.97\pm0.01$ (healthy controls) for \textit{Identification} of COVID-19 and $0.97\pm0.01$ Spearman Correlation for \textit{Severity quantification}. We release all the code and create a public leaderboard, where other community members can test their models on our test dataset.}




\end{abstract}

\end{frontmatter}


\input{content/1_introduction}

\input{content/2_method.tex}
\input{content/3_data.tex}
\input{content/4_experiments.tex}
\input{content/5_results.tex}
\input{content/6_discussion.tex}

\bibliography{main,triage,covid_algos,medical}

\end{document}

%% file: content/1_introduction.tex
\section{Introduction}
\label{sec:intro}

During the first months of 2020, COVID-19 infection spread worldwide and affected millions of people \citep{li2020early}. \answer{Though a virus-specific reverse transcription-polymerase chain reaction (RT-PCR) testing remains the gold standard \citep{worldclinical}, chest imaging, including computed tomography (CT), is helpful in diagnosis and patient management \citep{bernheim2020chest, akl2020use, rubin2020others}. Moreover, compared to RT-PCR, CT has higher sensitivity (98\% compared to 71\% at $p \le 0.001$) for some cohorts \cite{fang2020sensitivity}. Fleischner Society has addressed the role of thoracic imaging in COVID-19, providing recommendations intended to guide medical practitioners with one scenario including medical triage in moderate-to-severe clinical features and a high pretest probability of disease \citep{rubin2020others}. Radiology departments can respond to the pandemic by division into four areas (contaminated, semi-contaminated, buffer, and clean), strict disinfection and management criteria \citep{huang2020battle}. The International Society of Radiology surveyed current practices in managing patients with COVID-19 in 50 radiology departments representing 33 countries across all continents. In symptomatic patients with suspected COVID-19, imaging was performed in 89\% of cases, in 34\% of cases - chest CT. Faster results than molecular tests (51\%) and easy access (39\%) were the main reasons for imaging use \citep{blavzic2020use}.}

\begin{figure}[b!]
    \begin{center}
      \includegraphics[width=1.\linewidth]{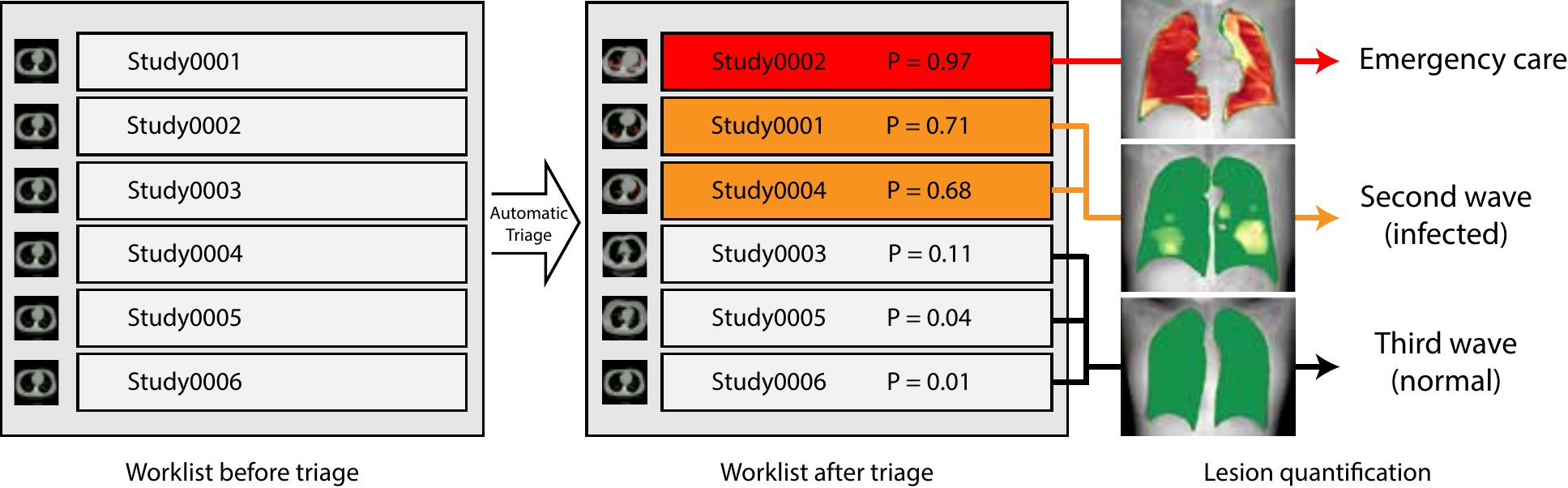}
      \caption{
      A schematic representation of the automatic triage process. Left: the chronological order of the studies. Center: re-prioritized order to highlight findings requiring radiologist's attention (\textit{P} denotes COVID-19 \textit{Identification} probability). Right: accompanying algorithm-generated X-ray-like series to assist the radiologist in fast decision making (color bar from green to red denotes \textit{Severity} of local COVID-19-related changes).
      }
      \label{fig:triage2}
    \end{center}
\end{figure}

The pandemic dramatically increased the need for medical care and resulted in the overloading of healthcare systems \citep{tanne2020covid}. Many classification and segmentation algorithms were developed to assist radiologists in COVID-19 identification and severity quantification, see Sec.~\ref{sec:related_covid}. However, little research has been conducted to investigate automatic image analysis for triage, i.e. ranking of CT studies. During an outbreak, many CT scans require rapid decision-making to sort patients into those who need care right now and those who will need scheduled care \citep{mei2020artificial}. Therefore, the study list triage is relevant and may shorten the report turnaround time by increasing the priority of CT scans with suspected pathology for faster interpretation by a radiologist compared to other studies, see Fig. \ref{fig:triage2}.

The triage differs from other medical image analysis tasks, as in this case, automatic programs provide the first reading. The radiologist then becomes the second reading. 
Technically, many of the developed methods may provide a priority score for triage, e.g., output probability of a classifier or the total lesion volume extracted from a binary segmentation mask. However, these scores must be properly used. We assume that there are two 
different triage problems:
\begin{enumerate}
    \item \textit{Identification}. The first challenging task is to identify studies of patients with COVID-19 and prioritize them so the physician can isolate potentially infected patients as early as possible \citep{sverzellati2020integrated}.
    \item \textit{Severity quantification}. Second, within COVID-19 patients, a triage algorithm must prioritize those who will require emergency medical care \citep{kherad2020computed}.
\end{enumerate}

Binary classification provides a direct way to formalize \textit{Identification}, but the optimal computer science approach to estimate \textit{Severity} is not as obvious. 
It was shown that human-based quantitative analysis of chest CT helps assess the clinical severity of COVID-19. 
\citep{colombi2020well} had quantified affected pulmonary tissue and established a high correlation between the healthy pulmonary tissue volume and the outcomes (transfer to an intensive care unit or death). The threshold value for the volume of healthy pulmonary tissue was 73\%. This result and similar ones motivate clinical recommendations in several countries: COVID-19 patients need to be sorted based on quantitative evaluation of lung lesions. 

\answer{
In particular, the Russian Federation adopted the following approach \citep{morozov2020covid}: the volume ratio of lesions in each lung is calculated separately and the maximal ratio is treated as the overall \textbf{severity score}.
However, manual binary segmentation of the affected lung tissue 
is extremely time-consuming and may take several hours \citep{shan2020lung}. For this reason, a visual semi-quantitative scale was implemented rather than a fully quantitative one. The original continuous index is split up into five categories: from CT-0 to CT-4 with a 25\% step so that CT-0 corresponds to normal cohort and CT-4 - to 75\%-100\% of damaged lung tissue. Patients with CT-3 (severe pneumonia) are hospitalized, and CT-4 (critical pneumonia) are admitted to an intensive care unit.
The scale is based on a visual evaluation of approximate lesion volume in both lungs (regardless of postoperative changes). 

A retrospective study \citep{morozov2020-lethal-outcome} analyzed the CT 0-4 scores and lethal outcomes in 13003 COVID-19 patients. The trend of directional changes in the deceased patient proportion using CT 0-4 score demonstrated a statistically significant result ($p < 0.0001$). The chance of a lethal outcome increased from CT-0 to CT-4 by 38\% on the average (95\% CI 17.1-62.6\%). Another retrospective analysis \citep{petrikov-semiotics} found a significant correlation between an increase of CT grade and clinical condition deterioration ($r = 0.577$).

}

\begin{figure}[]
    \begin{center}
      \includegraphics[width=1.\linewidth]{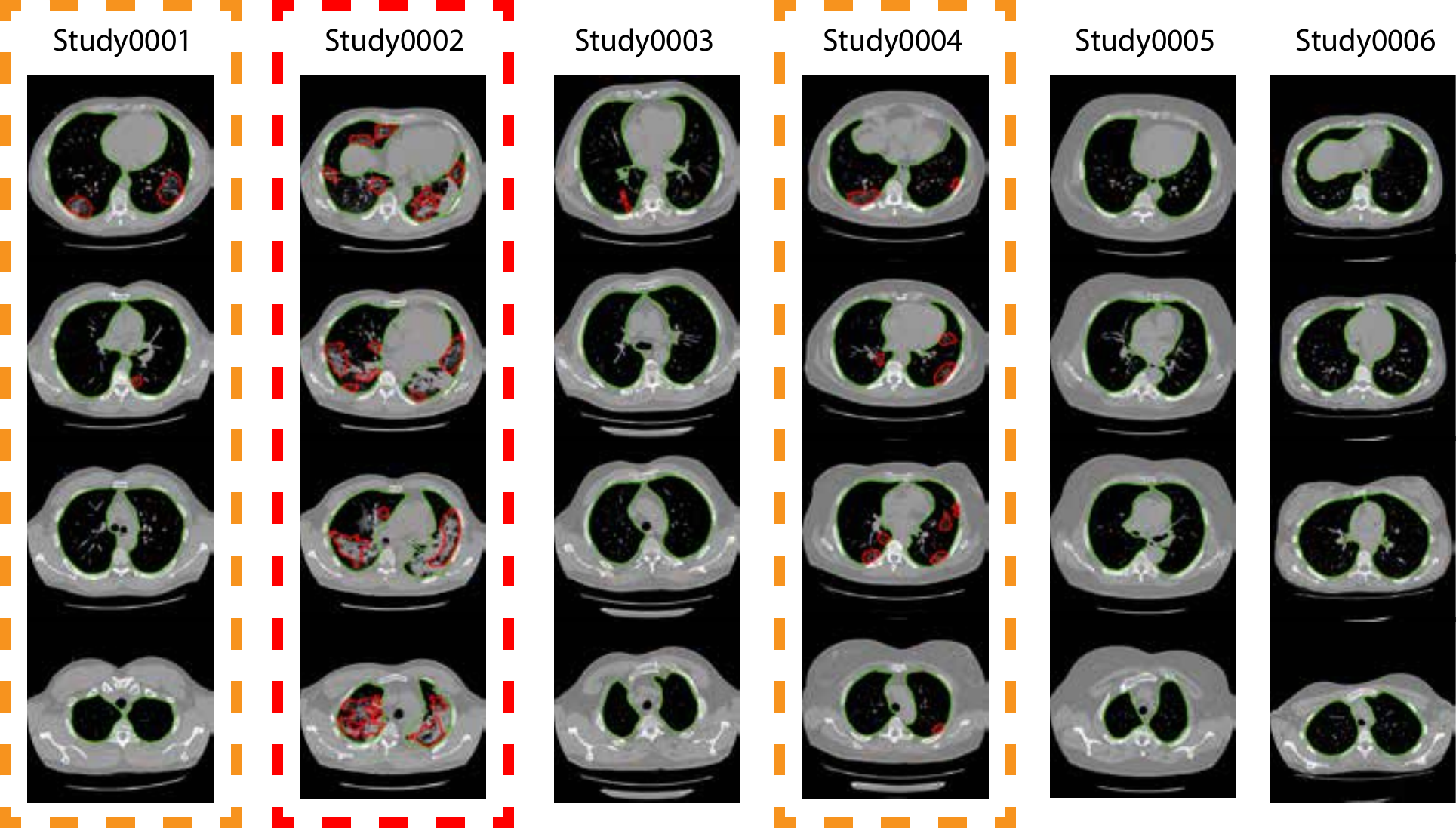}
      \caption{
      An example of joint COVID-19 identification and severity estimation by the proposed method for several studies.
      }
      \label{fig:studies}
    \end{center}
\end{figure}

These two triage strategies, \textit{Identification} and \textit{Severity quantification}, are not mutually exclusive, and their priority may change depending on the patient population structure and current epidemiological situation.
\begin{itemize}
    \item An outpatient hospital in an area with a small number of infected patients may rely on \textit{Identification} solely.
    \item An infectious diseases hospital may use \textit{Severity quantification} to predict the need for artificial pulmonary ventilation and intensive care units.
    \item Finally, an outpatient hospital during an outbreak needs both systems to identify and isolate COVID patients as well as quantify disease severity and route severe cases accordingly.
\end{itemize}
This paper explores the automation of both \textit{Identification} and \textit{Severity quantification} intending to create a robust system for all scenarios, see Fig. \ref{fig:studies}.

\subsection{Related work}

\subsubsection{CT analysis for COVID-19 Identification and Severity Estimation}
\label{sec:related_covid}

As briefly discussed above, there are two main machine learning-based approaches to analyze COVID-19 chest CTs: identification and severity quantification. In both cases, researchers usually calculate a continuous index of COVID-19 presence or severity (depending on their task). An overview of the existed indices can be found in Tab. \ref{tab:covid_alogs}. Below, we present only some of the existing CT-based algorithms for a more comprehensive review we refer to \citep{shi2020review}. 

\begin{table}[]
    \caption{Overview of continuous output indices proposed in previous works. The Type column denotes index type: COVID-19 \textit{identification}, COVID-19 \textit{severity} or \textit{both}.  Type of the Identification is given in brackets COVID vs : P - Pneumonia, NP - non-Pneumonia, HC - Healthy controls, N - Nodules, C - Cancer. The Metric column contains reported ROC AUC values unless otherwise indicated. \textbf{Remarks}. 1. Accuracy because ROC AUC was not reported. 2. The metric was provided for the identification problem only. 3. Pearson correlation. 4. The average volume error, measured in cm$^3$. 5. The paper does not provide an index, Dice score for the output masks is reported.}
    \vspace{0.2cm}
    \centering
    \resizebox{\textwidth}{!}{
    \begin{tabular}{lllr}
    \toprule
                                 Paper &                      Ranking index description &   Type & Metric \\
    \midrule
                      \cite{bai2020ai} &            Probabilities of 2.5D EfficientNet &  Iden. (P) &           .95 \\ 
              \cite{kang2020diagnosis} &           Probabilities of a NN for raidomics &       Iden. (P) &  Acc.$^1$ .96 \\ 
                   \cite{shi2020large} &             Probabilities of RF for radiomics &       Iden. (P) &           .94 \\ 
               \cite{li2020artificial} &                Probabilities of 2.5D ResNet50 &    Iden. (P, NP)&       .96 \\ 

                  \cite{wang2020prior} &         Probabilities of a 3D Resnet-based NN & Iden. (HC, P) &          .97 \\ 
                \cite{han2020accurate} &                     Probabilities of a 3D CNN &  Iden. (HC, P) &           .99 \\ 
                      \cite{jin2020ai} &                     Probabilities of ResNet50 & Iden. (HC, P, N) &           .99 \\ 
             \cite{jin2020development} &    Custom aggregation of a 2D CNN predicitons &       Iden. (HC, P) &           .97 \\ 
                 \cite{gozes2020rapid} &   Fractions of affected slices (by 2D ResNet) &       Iden. (HC, C) &           .99 \\ 
          \cite{amyar2020multi}        &      Probabilities of  3D Unet (encoder part) &       Iden. (HC, P)&           .97 \\ 
                 \cite{wang2020weakly} &                     Probabilities of a 3D CNN &Iden. (HC) &           .96 \\ 
                   \cite{chen2020deep} &           2D Bounding boxes + post-processing &       Iden. (other disease) &  Acc.$^1$ .99 \\ 

           \cite{gozes2020coronavirus} &          A score based on 2D ResNet attention &        Both (fever)&           .95$^2$ \\ 
     \cite{chaganti2020quantification} &   Affected lung percentage, a combined score  &        Sev. &      Corr.$^3$ .95 \\
                \cite{huang2020serial} &           Affected lung percentage by 2D Unet &        Sev. &             N/A \\
           \cite{shen2020quantitative} &  Affected lung percentage by non trainable CV &        Sev. &     Corr.$^3$ .81 \\
                   \cite{shan2020lung} &             Volume of segm. masks by a 3D CNN &        Sev. &     Vol.$^4$ 10.7 \\
                     \cite{fan2020inf} &                            Segmentation mask &         Sev. &      Dice$^5$ .60 \\
               \cite{tang2020severity} &                  Random Forrest probabilities &        Sev. &            .91 \\
    \bottomrule
    \end{tabular}
    }
    
    \label{tab:covid_alogs}
\end{table}
The majority of reviewed works use a pre-trained network for lung extraction or bounding box estimation as a necessary preprocessing step. We will skip the description of this step below for all works. 

\paragraph{Identification} Researchers usually treat this problem as binary classification, e.g. COVID-19 versus all other studies.  Likely, the most direct way to classify CT images with varying slice thicknesses is to train well established 2D convolutional neural networks. For example, authors of \citep{jin2020ai} trained ResNet-50 \citep{he2016deep} to classify images using the obtained lung mask. An interesting and interpretable way to aggregate slice predictions into whole-study predictions was proposed in \citep{gozes2020rapid}, where the number of affected slices was used as the final output of the model. Also, this work employed Grad-cam \citep{selvaraju2017grad} to visualize network attention. A custom slice-level prediction aggregation was proposed in \citep{jin2020development} to filter out false positives. 

The need for a post-training aggregation of slice prediction can be avoided by using 3D convolutional networks, \citep{han2020accurate,wang2020weakly}. 
\citep{wang2020prior} proposed a two-headed architecture based on 3D-ResNet. This approach is a way to obtain hierarchical classification as the first head was trained to classify CTs with and without pneumonia. In contrast, the second one aimed to distinguish COVID-19 from other types of pneumonia. 
Alternatively, slice aggregation may be inserted into network architectures to obtain an end-to-end pipeline, as was proposed in \citep{li2020artificial, bai2020ai}. Within this setup, all slices are processed by a 2D backbone (ResNet50 for \citep{li2020artificial}, EfficientNet \citep{tan2019efficientnet} for \citep{bai2020ai}) while the final classification layers operate with a pooled version of feature maps from all slices.  \answer{ \cite{amyar2020multi} proposed a multi-head architecture to solve both Segmentation and Identification problems in an end-to-end manner. They have used a 3D Unet backbone with an additional classification head after the encoder part. Even though they did not tackle the problem of severity identification, they demonstrate that solving two tasks jointly could benefit both.}

\paragraph{Segmentation} The majority of papers for tackling severity estimation are segmentation based. For example, the total absolute volume of involved lung parenchyma can be used as a quantitative index \citep{shan2020lung}. Relative volume (i.e., normalized by the total lung volume) is a more robust approach taking into account the normal variation of lung sizes. Affected lung percentage was estimated in several ways including a non-trainable Computer Vision algorithm \citep{shen2020quantitative}, 2D Unet \citep{huang2020serial}, and 3D Unet \citep{chaganti2020quantification}. Alternatively, an algorithm may predict the severity directly, e.g., with Random Forrest based on a set of radiomics features \citep{tang2020severity} or a neural network.

As discussed above, many papers address either COVID-19 identification or severity estimation. However, little research has been conducted to study both tasks simultaneously. \citep{gozes2020coronavirus} proposed an original Grad-cam-based approach to calculate a single attention-based score. Though the authors mentioned both identification and severity quantification in the papers, they do not provide direct quality metrics for the latter. 

\subsubsection{Deep learning for triage}
\label{sec:related_triage}

As mentioned above, we consider triage to be the process of ordering studies to be examined by a radiologist. There are two major scenarios where such an approach can be useful:
\begin{itemize}
\item Studies with a high probability of dangerous findings must be prioritized. The most important example is triage within emergency departments, where minutes of acceleration may save lives \citep{faita2020deep}, but it may be useful for other departments as well. For example, the study \citep{annarumma2019automated} estimates the average reporting delay in chest radiographs as 11.2 days for critical imaging findings and 7.6 days for urgent imaging findings. 

\item The majority of studies do not contain critical findings. This is a common situation for screening programs, e.g., CT-based lung cancer screening \citep{national2011national}. In this scenario, triage systems aim to exclude studies with the smallest probability of important findings to reduce radiologists' workload. 
\end{itemize}
 
Medical imaging may provide detailed information useful for automatic patient triage, as shown in several studies. \citep{annarumma2019automated} developed a deep learning-based algorithm to estimate the urgency of imaging findings on adult chest radiographs. The dataset included $470 388$ studies annotated in an automated way via text report mining. The Inception v3 architecture \citep{szegedy2016rethinking} was used to model clinical priority as ordinal data via solving several binary classification problems as proposed in \citep{lin2012reduction}. The average reporting delay was reduced to 2.7 and 4.1 days for critical and urgent imaging findings correspondingly in a simulation on historical data.  

A triage system for screening mammograms, another 2D image modality, was developed in \citep{yala2019deep}. The authors draw attention to reducing the radiologist's load by maximizing system recall. The underlying architecture is ResNet-18 \citep{he2016deep}, which was trained on $223 109$ screening mammograms. The model achieved 0.82 ROC AUC on the whole test population and demonstrated the capability to reduce workload by 20\%  while preserving the same level of diagnostic accuracy.

Prior research confirms that deep learning may assist in triage of more complex images such as volumetric CT. A deep learning-based system for rapid diagnosis of acute neurological conditions caused by stroke or traumatic brain injury was proposed in \citep{titano2018automated}. A 3D adaption of ResNet-50 \citep{korolev2017residual} analyzed head CT images to predict critical findings. To train the model, the authors utilized $37 236$ studies; labels were also generated by text reports mining.
The classifier's output probabilities served as ranks for triage, and the system achieved ROC AUC 0.73-0.88. Stronger supervision was investigated in \citep{chang2018hybrid}, where authors used 3D masks of all hemorrhage subtypes of $10 159$ non-contrast CT. The detection and quantification of 5 subtypes of hemorrhages were based on a modified Mask R-CNN \citep{he2017mask} extended by pyramid pooling to map 3D input to 2D feature maps \citep{lin2017feature}. More detailed and informative labels combined with an accurately designed method provide reliable performance as ROC AUC varies from 0.85 to 0.99 depending on hemorrhage type and size.
A similar finding was reported in \citep{de2018clinically} for optical coherence tomography (OCT). The authors employed a two-stage approach. First, 3D-Unet \citep{cciccek20163d} was trained on $877$ studies with dense 21-class segmentation masks. Then output maps for another $14 884$ cases were processed by a 3D version of DenseNet \citep{huang2017densely} to identify urgent cases. The obtained combination of two networks provided excellent performance achieving 0.99 ROC AUC.

\subsection{Contribution} 

\textit{First}, we highlighted the need for triage systems of two types for COVID-19 identification and severity quantification. We studied existing approaches and demonstrated that a system trained for one task shows low performance in the other. 
\textit{Second}, we developed a multitask learning-based approach to create a triage system for both types of problems, which achieves top results in both tasks. 
\textit{Finally}, we provided a framework for reproducible comparison of various models (see the details below). 

\subsubsection{Reproducible research}
A critical meta-review \citep{wynants2020systematic} of machine learning models for COVID-19 diagnosis highlights low reliability and high risk of biased results for all 27 reviewed papers, mostly due to a non-representative selection of control patients and poor analysis of results, including possible model overfitting.
The authors used \citep{wolff2019probast} PROBAST (Prediction model Risk Of Bias Assessment Tool), a systematic approach to validate the performance of machine learning-based approaches in medicine and identified the following issues.
\begin{enumerate}
    \item Poor patient structure of the validation set, including several studies where control studies were sampled from different populations.
    \item Unreliable annotation protocol where only one rater assessed each study without subsequent quality control or the model output influenced annotation.
    \item Lack of comparison with other well-established methods for similar tasks.
    \item Low reproducibility due to several factors such as unclear model description and incorrect validation approaches (e.g., slice-level prediction rather than study-level prediction).
\end{enumerate}
The authors concluded the paper with a call to share data and code to develop an established system for validating and comparing different models collaboratively.   

Though \citep{wynants2020systematic} is an early review and does not include many properly peer-reviewed papers mentioned above, we agree that current algorithmic research lacks reproducibility. Indeed, only one paper \citep{fan2020inf} among $18$ papers from section \ref{sec:related_covid} used open datasets for training and released the code. We aim to follow the best practices of reproducible research and address these issues in the following way.
\begin{enumerate}
    \item We selected fully independent test dataset and retrieved all COVID-19 positive and COVID-19 negative cases from the same population and the same healthcare system, see details in Sec. \ref{ssec:data:test}.
    \item Two raters annotated the test data independently. If raters contours are not aligned, the meta-rater requested annotation correction, see Sec. \ref{ssec:data:test}. 
  \item We carefully selected several key ideas from the reviewed works and implemented them within the same pipeline as our method, see Sec. \ref{sec:method}.
  \item We publicly release the code to share technical details of the compared architectures\footnote{https://github.com/neuro-ml/covid-19-multitask}. 
\end{enumerate}
Finally, we used solely open datasets for training; the test set includes some private images. We have also established an online leaderboard\footnote{https://github.com/neuro-ml/covid-19-multitask/wiki/Leaderboard} to compare different segmentation and classification methods on our test data.

%% file: content/2_method.tex
\section{Method}
\label{sec:method}

\answer{As discussed in Sec.~\ref{sec:intro} method should solve two tasks: identification of COVID-19 cases and ranking them in descending order of severity. Therefore, we structure Sec.~\ref{sec:method} as follows.
\begin{itemize}
    \item In Sec.~\ref{ssec:method:lungs} we describe lungs segmentation as a common preprocessing step for all methods.
    \item In Sec.~\ref{ssec:method:severity} we tackle the severity quantification task. We describe methods which predict segmentation mask of lesions caused by COVID-19 and provide a severity score based on that.
    \item In Sec.~\ref{ssec:method:identification} we discuss two straightforward baselines for the identification task. First is to use segmentation results and identify patients with non-empty lesions masks as COVID-19 positive. Second is to use separate neural network for classification of patients into COVID-19 positive or negative. However, as we show in Sec.~\ref{sec:results} these methods yield poor identification quality, especially due to false positive alarms in patients with bacterial pneumonia. 
    \item In Sec.~\ref{ssec:method:multitask} we propose a multitask model which achieves better COVID-19 identification results than the baselines. In particular, as we show in Sec.~\ref{sec:results}, this model successfully distinguishes between COVID-19 and bacterial pneumonia cases.
    \item In Sec.~\ref{ssec:method:metrics} we introduce quality metrics for both identification and severity quantification tasks to formalize the comparison of the methods.
\end{itemize}}


\subsection{Lungs segmentation}
\label{ssec:method:lungs}
We segment lungs in two steps. First, we predict single binary mask for both lungs \answer{including pathological findings, e.g. ground-glass opacity, consolidation, nodules and pleural effusion.} Then we split the obtained mask into separate left and right lungs' masks.
Binary segmentation is performed via fully-convolutional neural network in a standard fashion. Details of the architecture and training setup are given in Section \ref{ssec:exp:lungs}.

On the second step voxels within the lungs are clustered using $k$-means algorithm ($k=2$) with Euclidean distance as a metric between voxels. \answer{Then we treat resulting clusters as separate lungs.}

\subsection{COVID-19 severity quantification}
\label{ssec:method:severity}
\answer{To quantify COVID-19 severity we solve COVID-19-specific lesions segmentation task. Using predicted lungs' and lesions' masks, we calculate the lesions' to lung's volume ratio for each lung and use the maximum of two ratios as a final severity score for triage, according to recommendations discussed in Sec.~\ref{sec:intro}.}

\paragraph{Threshold-based}
\label{par:method:severity:cv}
As a baseline for lesions segmentation, we choose the thresholding-based method (since pathological tissues are denser than healthy ones, corresponding CT voxels have greater intensities in Hounsfield Units). The method consists of three steps. The first step implements thresholding: voxels with intensity value between $\text{HU}_{\text{min}}$ and $\text{HU}_{\text{max}}$ within the lung mask are assigned to the positive class. At the second step, we apply Gaussian blur with smoothing parameter $\sigma$ to the resulting binary mask and reassign the positive class to voxels with values greater than $0.5$. Finally, we remove 3D binary connected components with volumes smaller than $V_{\text{min}}$. The hyperparameters $\text{HU}_{\text{min}} = -700$, $\text{HU}_{\text{max}} = 300$, $\sigma = 4$ and $V_{\text{min}} = 0.1\%$ are chosen via a grid-search in order to maximize the average Dice score between predicted and ground truth lesions masks for series from training dataset.

\paragraph{U-Net}
\label{par:method:severity:unet}
The de facto standard approach for medical image segmentation is the U-Net model \citep{unet}. We trained two U-Net-based architectures for lung parenchyma involvement segmentation which we refer to as \textit{2D U-Net} and \textit{3D U-Net}. \textit{2D U-Net} independently processes the axial slices of the input 3D series. \answer{\textit{3D U-Net} processes 3D sub-patches of size $160 \times 160 \times 160$ and then stacks predictions for individual sub-patches to obtain prediction for the whole input 3D series. Thus, we do not need to downsample the input image under the GPU memory restrictions.}
For each model, we replace plain 2D and 3D convolutional layers with 2D and 3D residual convolutional blocks \citep{resnet}, correspondingly. Both models were trained using the standard binary cross-entropy loss (see other details in Sec. \ref{ssec:exp:segmentation}). 




\subsection{COVID-19 Identification}
\label{ssec:method:identification}
\answer{We formalize COVID-19 identification task as a binary classification of 3D CT series. CT series of patients with verified COVID-19 are positive class. CT series of patients with other lung diseases, e.g. bacterial pneumonia, non-small cell lung cancer, etc., as well as normal patients are negative class.}

\paragraph{Segmentation-based} One possible approach is to base the decision rule on the segmentation results: classify a series as positive if the segmentation-based severity score exceeds some threshold. \answer{We show that this leads to a trade-off between severity quantification and identification qualities: models which yield the best ranking results perform worse in terms of classification, and vice versa. Moreover, despite some segmentation-based methods accurately classify COVID-19 positive and normal cases, all of them yields a significant number of false positives in patients with bacterial pneumonia (see Sec.~\ref{ssec:results:seg}).}

\paragraph{ResNet50} Another approach is to tackle the classification task separately from segmentation and explicitly predict the probability that a given series \answer{is COVID-19 positive}. The advantage of this strategy is that we only need weak labels for model training, which are much more available than ground truth segmentations. 

To assess the performance of this approach we follow the authors of \citep{li2020artificial, bai2020ai} and train the ResNet50 \citep{resnet} which takes a series of axial slices as input and independently extracts feature vectors for each slice. \answer{After that the feature vectors are combined via a pyramid max-pooling operation \citep{pyramid} along all the slices. The resulting vector is passed into two fully connected layers followed by sigmoid activation which predicts the final COVID-19 probability for the whole series. In our paper, we denote this architecture as \textit{ResNet50}(see other details in Section \ref{ssec:exp:identification})}.

\subsection{Multitask} 
\label{ssec:method:multitask}

\begin{figure}[t]
    \begin{center}
      \includegraphics[width=1.\linewidth]{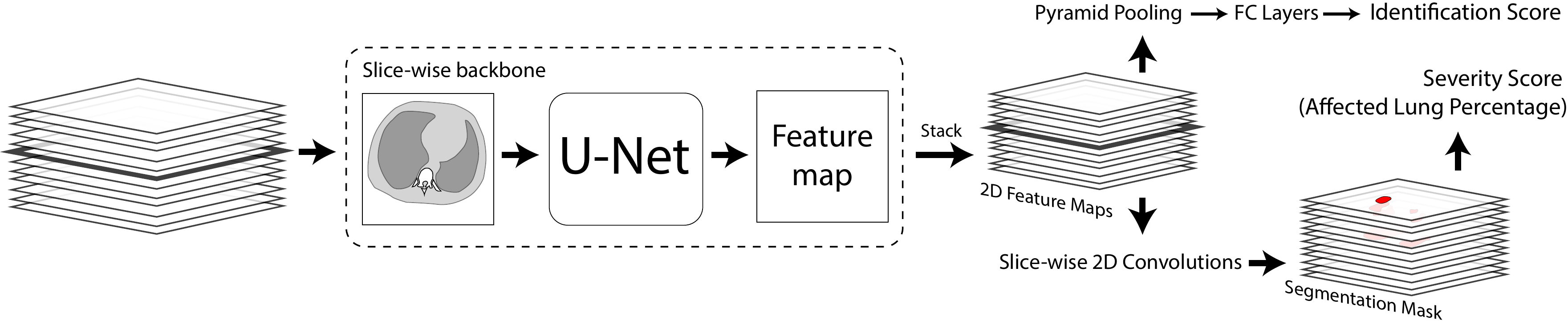}
      \caption{
      Schematic representation of the \textit{Multitask-Spatial-$1$} model. \textit{Identification} score is the probability of being a COVID-19 positive series; \textit{Severity} score is calculated using predicted lesions' mask and precomputed lungs' masks.
      }
      \label{fig:model}
    \end{center}
\end{figure}

\answer{Baselines for the identification task described in Sec~\ref{ssec:method:identification} do not perform well, as we show in Sec.~\ref{sec:results}. Therefore, we propose to solve the identification task simultaneously with the segmentation task via a single two-headed convolutional neural network.

The segmentation part of the architecture is slice-wise \textit{2D U-Net} model. As earlier, its output is used for the evaluation of the severity score.

The classification head shares a common intermediate feature map (per slice) with the segmentation part. These feature maps are stacked and aggregated into a feature vector via a pyramid pooling layer \citep{pyramid}. Finally, two fully connected layers followed by sigmoid activation transform the feature vector to the COVID-19 probability. 

Following \citep{amyar2020multi}, the shared feature maps can be the outputs of the U-Net's encoder and have no explicit spatial structure in the axial plane. We refer to this approach as \textit{Multitask-Latent}. In contrast, we argue that the identification task is connected to the segmentation task and the classification model can benefit from the spatial structure of the input features. Therefore, we propose to share the feature map from the very end of the U-Net architecture, as shown in Fig.~\ref{fig:model}. We refer to the resulting architecture as \textit{Multitask-Spatial-1}. More generally, shared feature maps can be taken from the $l$-th upper level of the U-Net's decoder. Together they form a 3D spatial feature map, which is aligned with the input 3D series downsampled in the axial plane by a factor of $2^{l - 1}$. We denote this approach as \textit{Multitask-Spatial-$l$}. Since \textit{2D U-Net} architecture has $7$ levels, $l$ can vary from $1$ to $7$.}



As a loss function we optimize a weighted combination of binary cross entropies for segmentation and classification (see other details in Section~\ref{sec:exp}). 

\answer{
\subsection{Metrics}
\label{ssec:method:metrics}
To assess the quality of classification of patients into positive, i.e. infected by COVID-19, and negative, i.e. with other lung pathologies or normal, we use areas under the ROC-curves (ROC-AUC) calculated on several subsamples of the test sample described in Sec.~\ref{ssec:data:test}.
\begin{itemize}
    \item The first subsample contains only COVID-19 positive and healthy subjects, while studies with other pathological findings are excluded (ROC-AUC COVID-19 vs Normal).
    \item The second subsample contains only patients infected by COVID-19 or bacterial pneumonia (ROC-AUC COVID-19 vs Bac. Pneum.).
    \item The third subsample contains COVID-19 positive patients and patients with lung nodules typical for non-small cell lung cancer (ROC-AUC COVID-19 vs Nodules).
    \item The last ROC-AUC is calculated on the whole test sample (ROC AUC COVID-19 vs All others).
\end{itemize}
ROC-curves are obtained by thresholding the predicted probabilities for ResNet50 and multitask models, and by thresholding the predicted severity score for segmentation-based methods.

We evaluate the quality of ranking studies in order of descending COVID-19 severity on the test subsample, which contains only COVID-19 positive patients. As a quality metric, we use Spearman's rank correlation coefficient (Spearman's $\rho$) between the severity scores $\mathbf{y}^\text{true}$ calculated for ground truth segmentations and the predicted severity scores $\mathbf{y}^{\text{pred}}$. It is defined as
$$
\rho(\mathbf{y}^\text{true}, \mathbf{y}^{\text{pred}}) = \frac{\text{cov}(\text{rg}(\mathbf{y}^\text{true}), \text{rg}(\mathbf{y}^{\text{pred}}))}{\sigma(\text{rg}(\mathbf{y}^{\text{true}})) \cdot \sigma(\text{rg}(\mathbf{y}^{\text{pred}}))},
$$
where $\text{cov}(\cdot, \cdot)$ is a sample covariance, $\sigma(\cdot)$ is a sample standard deviation and $\text{rg}(\mathbf{y})$ is the vector of ranks, i.e. resulting indices of $\mathbf{y}$ elements after their sorting in the descending order. 

To evaluate the COVID-19 lesions segmentation quality we use Dice score coefficient between the predicted and the ground truth segmentation masks. Similar to Spearman's $\rho$, we evaluate the mean Dice score only for COVID-19 positive cases.
}

%% file: content/3_data.tex
\section{Data}
\label{sec:data}
We used several public datasets in our experiments:
\begin{itemize}
    \item NSCLC-Radiomics and LUNA16 to create a robust lung segmentation algorithm.
    \item Mosmed-1110, MedSeg-29 and NSCLC-Radiomics to train and validate all the triage  models. 
    \item Mosmed-Test as a hold-out test set to assess the final performance of all the methods.
\end{itemize}

\subsection{Mosmed-1110}
\label{ssec:data:1110}
1110 CT scans from Moscow outpatient clinics were collected from 1st of March, 2020 to 25th of April, 2020, within the framework of outpatient computed tomography centers in Moscow, Russia \citep{morozov2020mosmeddata}.

Scans were performed on \textit{Canon (Toshiba) Aquilion 64} units in with standard scanner protocols and, particularly 0.8 mm inter-slice distance. However, the public version of the dataset contains every 10th slice of the original study, so the effective inter-slice distance is 8mm.

The quantification of COVID-19 severity in CT was performed with the visual semi-quantitative scale adopted in the Russian Federation and Moscow in particular \citep{morozov2020covid}. According to this grading, the dataset contains 254 images without COVID-19 symptoms. The rest is split into 4 categories: CT1 (affected lung percentage 25\% or below, 684 images), CT2 (from 25\% to 50\%, 125 images), CT3 (from 50\% to 75\%, 45 images), CT4 (75\% and above, 2 images).

Radiologists performed an initial reading of CT scans in clinics, after which experts from the advisory department of the Center for Diagnostics and Telemedicine (CDT) independently conducted the second reading as a part of a total audit targeting all CT studies with suspected COVID-19. 

Additionally, 50 CT scans were annotated with binary masks depicting regions of interest (ground-glass opacities and consolidation).

\subsection{MedSeg-29}
MedSeg web-site\footnote{\url{https://medicalsegmentation.com/covid19/}} 
shares $2$ publicly available datasets of annotated volumetric CT images. 
The first dataset consists of $9$ volumetric CT scans from a web-site\footnote{\url{https://radiopaedia.org/articles/covid-19-3}} 
that were converted from JPG to Nifti format.
The annotations of this dataset include lung masks and COVID-19 masks segmented by a radiologist. The second dataset consists of $20$ volumetric CT scans shared by \citep{external20}. The left and rights lungs, and infections are labeled by two radiologists and verified by an experienced radiologist.


\subsection{NSCLC-Radiomics}
\label{ssec:data:nsclc}
NSCLC-Radiomics dataset \citep{nsclc1, nsclc2} represents a subset of The Cancer Imaging Archive NSCLC Radiomics collection \citep{tcia}. It contains left and right lungs segmentations annotated on 3D thoracic CT series of 402 patients with diseased lungs. Pathologies -- lung cancerous nodules, atelectasis and pleural effusion -- are included in the lung volume masks. Pleural effusion and cancerous nodules are also delineated separately, when present.

\answer{Automatic approaches for lungs segmentation often perform inconsistently for patients with diseased lungs, while it is usually the main case of interest. Thus, we use NSCLC-Radiomics to create robust for pathological cases lungs segmentation algorithm. Other pathologies, e.g. pneumothorax, that are not presented in NSCLC-Radiomics could also lead to poor performance of lungs segmentation. But the appearance of such pathologies among COVID-19 cases is extremely rare. For instance, it is less than $1\%$ for pneumothorax \citep{zantah2020pneumothorax}. Therefore, we ignore the possible presence of other pathology cases, while training and evaluating our algorithm.}


\subsection{LUNA16}
LUNA16 \citep{luna16} is a public dataset for cancerous lung nodules segmentation. It includes $888$ annotated 3D thoracic CT scans from the LIDC/IDRI database \citep{armato2011lung}. Scans widely differ by scanner manufacturers (17 scanner models), slice thicknesses (from $0.6$ to $5.0$ mm), in-plane pixel resolution (from $0.461$ to $0.977$ mm), and other parameters. Annotations for every image contain binary masks for the left and right lungs, the trachea and main stem bronchi, and the cancerous nodules. The lung and trachea masks were originally obtained using an automatic algorithm \citep{van2009automatic} and the lung nodules were annotated by $4$ radiologists \citep{armato2011lung}. We also excluded 7 cases with absent or completely broken lung masks and extremely noisy scans. 

\subsection{Mosmed-Test}
\answer{
We created a test set as a combination of public and private datasets to ensure the following properties}
\begin{itemize}
\item \answer{All cases are full series without missing slices and/or lacking metadata fields (e.g., information about original Hounsfield units)}.
\item \answer{Data for all classes comes from the same population and the same healthcare system to avoid domain shifts within test data}.
\end{itemize}

\label{ssec:data:test}

\paragraph{COVID-19 positive} \answer{It is a subsample \citep{mosmed20} of $42$ CT studies collected from 20 patients in an infectious diseases hospital during the second half of February 2020, at the beginning of the Russian outbreak. We removed $5$ cases with artifacts related to patients' movements while scanning. The remaining $37$ cases were independently assessed by two raters (radiologists with 2 and 5 years of experience) who have annotated regions of interest (ground-glass opacities and consolidation) via MedSeg\footnote{\url{https://www.medseg.ai/}} annotation tool for every of the $37$ Mosmed-Test series. 

During this annotation process, we identified that 5 cases have no radiological signs of COVID-19, so we removed these studies from the list of COVID-19 positives. We iteratively verified annotations based on two factors: Dice Score between two rates and missing (by one of the raters) large connected components of the mask. The discrepancy between the two raters has been analyzed until we reached a consensus -- $0.87 \pm 0.17$ Dice Score over $32$ COVID-19 infected cases. 

It's also important to note that \citep{mosmed20} data were collected at inpatient clinics, whereas Mosmed-1110 is a subset of Moscow out-patient clinics database created from two to six weeks later, which guarantees that studies are not duplicated.}

\paragraph{Bacterial pneumonia} \answer{We randomly selected 30 studies from a list of patients with radiological signs of community-acquired bacterial pneumonia; images were acquired in 2019. This database is a private one as we do not aware of public bacterial pneumonia data coming from this population.} 

\paragraph{Lung nodules} \answer{We use a subset of a public dataset \citep{mosmed500} containing 500 chests CT scans randomly selected from patients over 50 years of age. We selected 30 cases randomly among cases with radiologically verified lung nodules. }


\paragraph{Normal controls} \answer{This subset consists of two parts: 5 cases mentioned above from  \citep{mosmed20} without radiological signs of COVID-19 and 26 studies \citep{mosmed500} without lung nodules larger than 5mm and other lung pathologies.}


%% file: content/4_experiments.tex
\section{Experiments}
\label{sec:exp}
\answer{
We design our experiments in order to objectively compare all the triage models described in Sec.~\ref{sec:method}. For that purposes all the methods are evaluated using the mean values and the standard deviations of the same quality metrics defined in Sec.~\ref{ssec:method:metrics} on the same hold-out test dataset described in Sec.~\ref{ssec:data:test}. We believe, that the experimental design for training neural networks for triage described in Sec.~\ref{ssec:exp:segmentation} and \ref{ssec:exp:identification} exclude overfitting.
}

All computational experiments were conducted on Zhores supercomputer \cite{zacharov2019zhores}.

\subsection{Preprocessing}
In all our experiments we use the same preprocessing applied separately for each axial slice: rescaling to a pixel spacing of $2\times2$mm and intensity normalization to the $[0, 1]$ range.

In our COVID-19 identification and segmentation experiments we crop the input series to the bounding box of the lungs' mask predicted by our lungs segmentation network.

\answer{We further show (Sec.~\ref{sec:results}) that this preprocessing is sufficient for all models. Despite the diversity of the training dataset, all models successfully adapt to the test dataset.}

\subsection{Lungs segmentation}
\label{ssec:exp:lungs}
For the lungs segmentation task we choose a basic U-Net \citep{unet} architecture with 2D convolutional layers, individually apply to each axial slice of an incoming series. The model was trained on NSCLC-Radiomics and LUNA16 datasets for 16k batches of size 30. We use Adam \citep{adam} optimizer with default parameters and an initial learning rate of 0.001, which was decreased to 0.0001 after 8k batches.

\answer{We assess the model's performance using 3-fold cross-validation and additionally using MedSeg-29 dataset as hold-out set. Dice Score of cross-validation is $0.976 \pm 0.023$ for both LUNA16 and NSCLC-Radiomics datasets, and $0.962 \pm 0.023$ only on NSCLC-Radiomics dataset. The latter result confirms our model to be robust to the cases with pleural effusion. 
Dice Score on MedSeg-29 dataset is $0.976 \pm 0.013$, which shows the robustness of our model to the COVID-19 cases.}

\answer{
\subsection{COVID-19 segmentation}
\label{ssec:exp:segmentation}
We use all the available 79 images of COVID-19 positive patients with annotated lesions masks (50 images from Mosmed-1110 and 29 images from MedSeg-29) to train the threshold-based, \textit{2D U-Net}, \textit{3D U-Net} models.

Additionally, we train the \textit{2D U-Net}'s architecture on the same 79 cases along with 402 images from the NSCLC-Radiomics dataset. These 402 images were acquired long before the COVID-19 pandemic, therefore we assume that ground truth segmentations for them are zero masks. During training this model we resample series such that batches contain approximately equal numbers of COVID-19 positive and negative cases. We refer to this model as \textit{2D U-Net+}. 

\textit{2D U-Net} and \textit{2D U-Net+} were trained for 15k batches using Adam \citep{adam} optimizer with default parameters and an initial learning rate of 0.0003. Each batch contains 5 series of axial slices. \textit{3D U-Net} was optimized via plain stochastic gradient descent for 10k batches using a learning rate of 0.01. Each batch consists of 16 3D patches.

In order to estimate mean values and standard deviations of models' quality metrics defined in Sec.~\ref{ssec:method:metrics} each segmentation network was trained 3 times with different random seeds. Resulting networks were evaluated on the hold-out test dataset, described in Sec.~\ref{ssec:data:test}.

\subsection{ResNet50 and multitask models}
\label{ssec:exp:identification}
The remaining 806 positive images without ground truth segmentations and 254 negative images from the Mosmed-1110 and 402 negative images from NSCLC-Radiomics were split 5 times in a stratified manner into a training set and a validation set. Each of the 5 validation sets contains 30 random images. 

For each split we train the \textit{ResNet50} and the classification heads of \textit{Multitask-Latent}, \textit{Multitask-Spatial-1} and \textit{Multitask-Spatial-4} models on the defined training set, while segmentation heads of the multitask models were trained on the same 79 images, as \textit{2D U-Net} (see Sec.~\ref{ssec:exp:segmentation}). 

For each network on each training epoch we evaluate the ROC-AUC between the predicted COVID-19 probabilities and the ground truth labels on the validation set. We save the networks' weights which resulted in the highest validation ROC-AUC during training. For all the multitask models as well as for \textit{ResNet50} top validation ROC-AUCs exceeded 0.9 for all splits.

All the networks were trained for 30k batches using Adam \citep{adam} optimizer with default parameters and an initial learning rate of 0.0003 reduced to 0.0001 after 24k batches. Each batch contained 5 series of axial slices.

During training the multitask models we resample examples such that batches contained an approximately equal number of examples which were used to penalize either classification or segmentation head. However, we multiplied by 0.1 the loss for the classification head, because it resulted in better validation ROC-AUCs.

For each of 5 splits, we evaluated each trained network on the hold-out test dataset described in Sec.~\ref{ssec:data:test}. We report the resulting mean values and standard deviations of the quality metrics in Sec.~\ref{sec:results}. 
}








%% file: content/5_results.tex
\section{Results}
\label{sec:results}

\answer{In this section we report and discuss the results of the experiments described in Sec.~\ref{sec:exp}. In Tab.~\ref{tab:results} we compare all the methods described in Sec.~\ref{sec:method} using quality metrics defined in Sec.~\ref{ssec:method:metrics} and evaluated on the test dataset described in Sec.~\ref{ssec:data:test}.}

\begin{table}[H]
    \centering
    \caption{Quantitative comparison of all the methods discussed in Section \ref{sec:method}. Trade-off between qualities of COVID-19 identification and ranking by severity is observed for segmentation-based methods. The proposed \textit{Multitask-Spatial-1} model yields the best identification results. Results are given as $mean \pm std$.}
    \resizebox{\textwidth}{!}{%
    \begin{tabular}{l c c c c c c}
        \toprule
        
        & \multicolumn{4}{c}{ROC-AUC (COVID-19 vs $\cdot$)} & \multirow{2}{*}{\shortstack{Spearman's $\rho$}} & \multirow{2}{*}{Dice Score} \\
        \cmidrule(lr){2-5}
        & vs All others & vs Normal & vs Bac. Pneum. & vs Nodules &  &  \\
        
        \midrule
        Thresholding & $.51 \pm .00$ & $.68 \pm .00$ & $.46 \pm .00$ & $.45 \pm .00$ & $.92 \pm .00$ & $.42 \pm .00$ \\
        
        \cmidrule{2-7}
        
        3D U-Net & $.76 \pm .02$ & $.89 \pm .02$ & $.59 \pm .01$ & $.79 \pm .03$ & $\mathbf{.97} \pm .01$ & $\mathbf{.65} \pm .00$ \\
        
        \cmidrule{2-7}
        
        2D U-Net & $.78 \pm .01$ & $.93 \pm .01$ & $.62 \pm .01$ & $.79 \pm .00$ & $\mathbf{.97} \pm .00$ & $.63 \pm .00$ \\
        
        \cmidrule{2-7}
        
        2D U-Net+ & $.86 \pm .01$ & $\mathbf{.98} \pm .01$ & $.68 \pm .02$ & $.91 \pm .01$ & $.80 \pm .03$ & $.59 \pm .01$ \\
        
        \cmidrule{2-7}
        
        ResNet50 & $.62 \pm .19$ & $.67 \pm .21$ & $.55 \pm .13$ & $.65 \pm .22$ & N/A & N/A \\
        
        \cmidrule{2-7}
        
        Multitask-Latent & $.79 \pm .06$ & $.84 \pm .05$ & $.73 \pm .06$ & $.80 \pm .07$ & $\mathbf{.97} \pm .00$ & $.61 \pm .02$ \\
        
        \cmidrule{2-7}
        
        Multitask-Spatial-4 &  $.89 \pm .03$ & $.94 \pm .03$ & $.83 \pm .05$ & $.91 \pm .03$ & $\mathbf{.98} \pm .00$ & $.61 \pm .02$ \\
        
        \cmidrule{2-7}
        
        Multitask-Spatial-1 & $\mathbf{.93} \pm .01$ & $\mathbf{.97} \pm .01$ & $\mathbf{.87} \pm .01$ & $\mathbf{.93} \pm .00$ & $\mathbf{.97} \pm .01$ & $.61 \pm .02$ \\
        
        \bottomrule
    \end{tabular}}
    \label{tab:results}
\end{table}

\subsection{Segmentation-based methods}
\label{ssec:results:seg}
In this subsection we discuss the performance of four methods: the threshold-based baseline, \textit{3D U-Net}, \textit{2D U-Net} and \textit{2D U-Net+}.

\begin{figure}
    \centering
    \vspace{-1cm}
    \includegraphics[width=.95\textwidth]{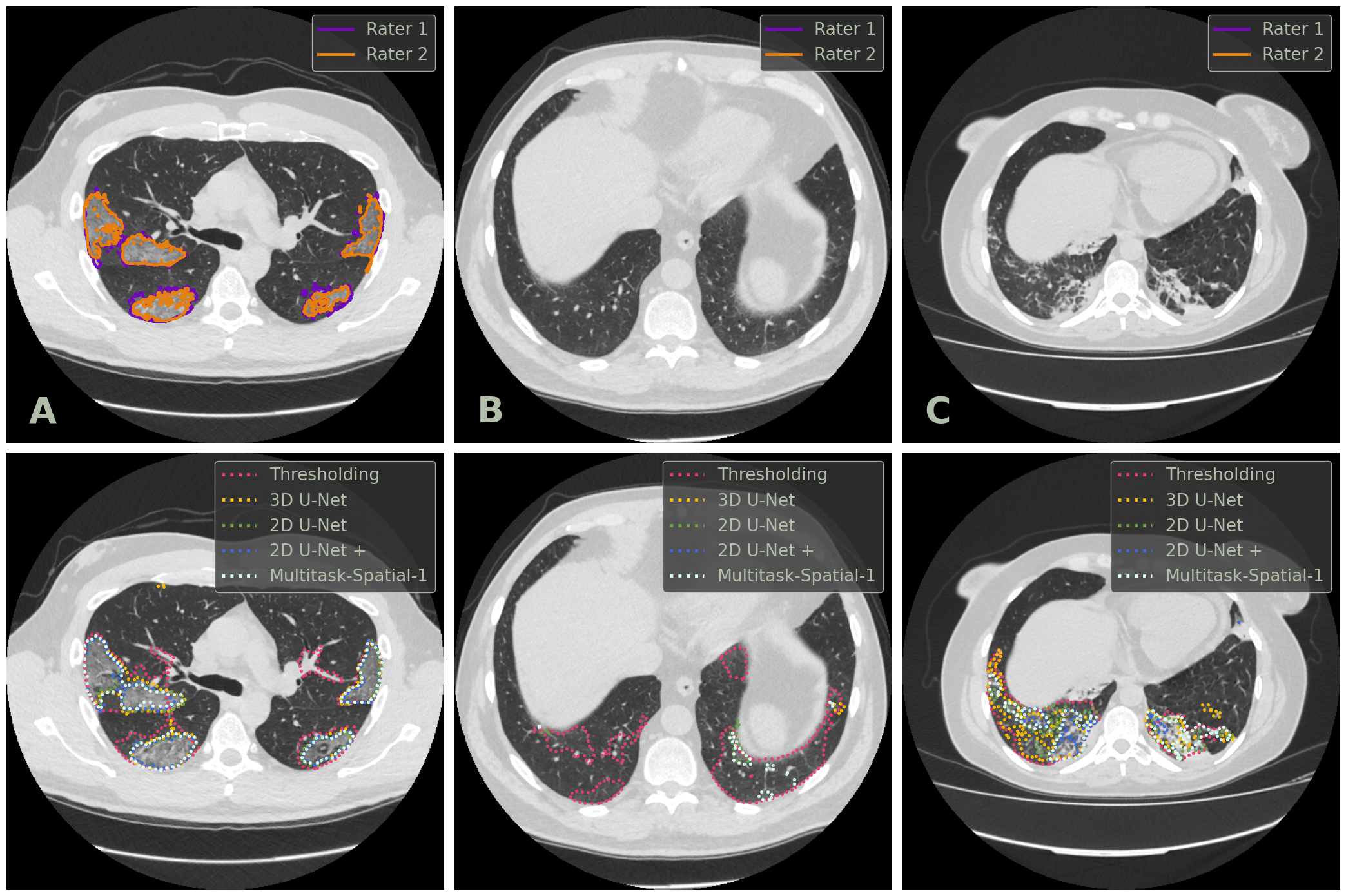}
    \caption{
        \answer{Examples of axial CT slices from the test dataset along with ground truth annotations (first row) and predicted masks (second row) of COVID-19-specific lesions. \textit{Column A}: COVID-19 positive case; \textit{Column B}: normal case; \textit{Column C}: case with bacterial pneumonia. Lesions' masks are represented by the contours of their borders for clarity.}
    }
    \label{fig:contours}
\end{figure}

We expect two major weaknesses of the threshold-based method: False Positive (FP) predictions on the vessels and bronchi, and inability to distinguish COVID-19 related lesions from other pathological findings. It is clearly seen from the extremely low ROC-AUC scores (Tab.~\ref{tab:results}). One could also notice massive FP predictions even in healthy cases (Fig. \ref{fig:contours}, column B). However, the method often provides a reasonable segmentation of the lesion area (Fig. \ref{fig:contours}, column A).

\answer{
Neural networks considerably outperform the threshold-based baseline in terms of any quality metric. We observe neither quantitative (Tab.~\ref{tab:results}) nor qualitative (Fig.~\ref{fig:contours}) significant difference between \textit{2D U-Net}'s and \textit{3D U-Net}'s performances. They yield accurate severity scores within the COVID-19 positive population (Spearman's $\rho = 0.97$). However, severity scores quantified for the whole test dataset do not allow to accurately distinguish between COVID-19 positive cases and cases with other pathological findings (ROC-AUC COVID-19 vs Bac. Pneum. $\approx 0.6$, ROC-AUC COVID-19 vs Nodules $=0.79$) due to FP segmentations (Fig.~\ref{fig:contours}, columns B and C).  

As one could expect, training on images with non-small cell lung cancer tumors from NSCLS-Radiomics dataset results in the enhancement of ROC-AUC vs Nodules (0.91 for \textit{2D U-Net+} compared to 0.79 for \textit{2D U-Net}). Interestingly, in this experiment we observe a degradation in terms of Spearman's $\rho$ for ranking of COVID-19 positive cases (0.8 for \textit{2D U-Net+} compared to 0.97 for \text{2D U-Net}). We conclude that one should account for this trade-off and use an appropriate training setup depending on the task.

All the segmentation-based models perform poorly in terms of classification into COVID-19 and bacterial pneumonia (ROC AUC COVID-19 vs Bac. Pneum. $\le 0.7$). This motivates to discuss the other methods.
}



\subsection{ResNet50}
Despite that validation ROC-AUCs for all the trained \textit{ResNet50} networks exceed 0.9, their performance on the test dataset is extremely unstable: ROC-AUC COVID-19 vs All varies from 0.43 to 0.85 (also see standard deviations in Tab.~\ref{tab:results}). 

\subsection{Multitask models}

\begin{figure}
\centering
    \includegraphics[width=1\textwidth]{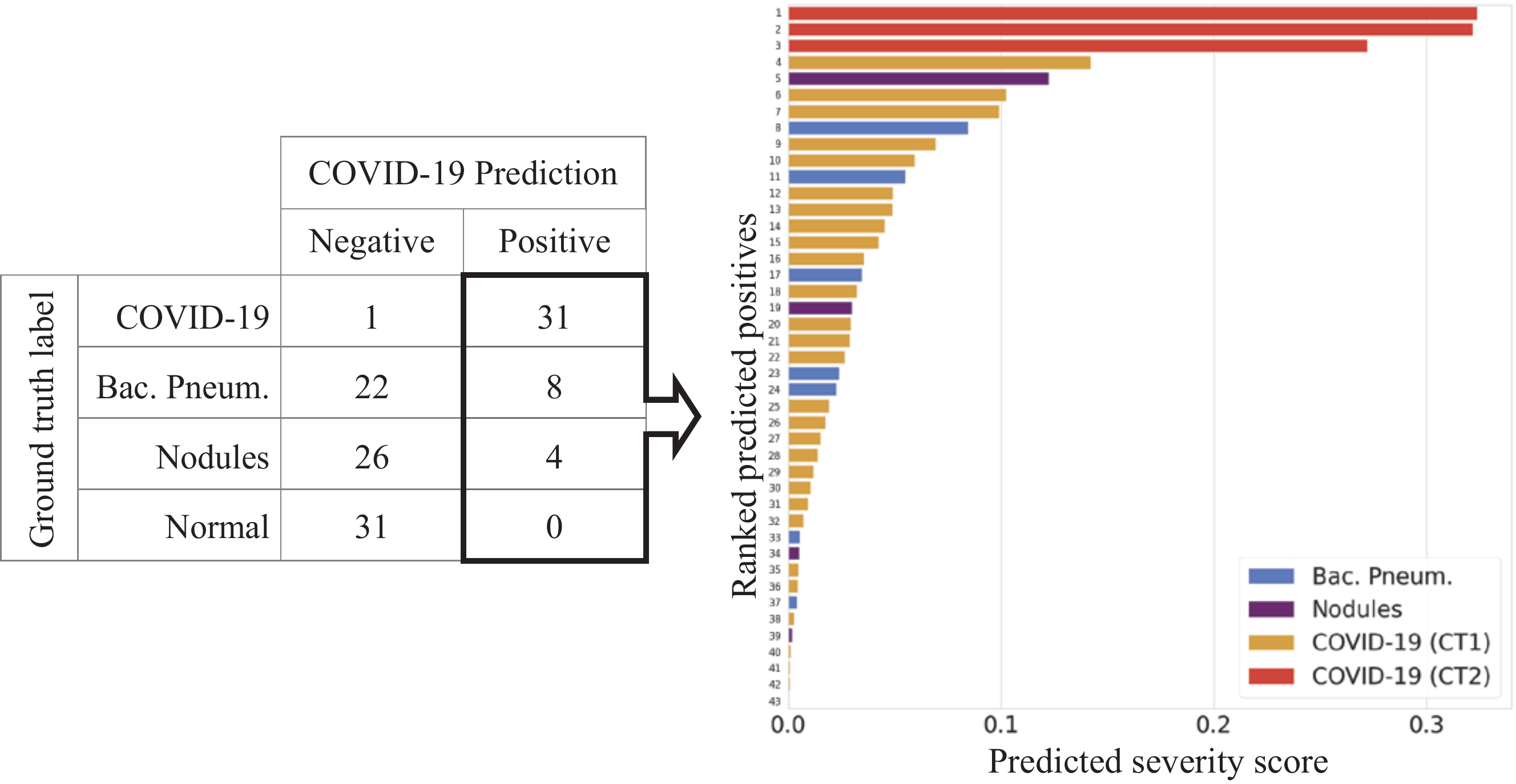}
    \caption{
        \answer{COVID-19 triage: identification of COVID-19 positive patients (left) and ranking them in the descending order of severity (right) via the proposed single \textit{Multitask-Spatial-1} model. In the right plot bars correspond to the ranked studies. Absolute values of the predicted affected lungs fractions are represented as bars' lengths along the $x$-axis. The bars' colors denote ground truth labels.}
    }
\label{fig:cls_rnk}
\end{figure}

\answer{
In this subsection we discuss the performance of \textit{Multitask-Latent}, \textit{Multitask-Spatial-4} and the proposed \textit{Multitask-Spatial-1} models on identification, segmentation and severity quantification tasks in comparison to each other, \textit{ResNet50} and segmentation-based methods.

As seen from mean values and standard deviations of ROC-AUC scores in Tab.~\ref{tab:results}, \textit{Multitask-Latent} model yields better and more stable identification results than \textit{ResNet50}. Both these models classify the latent representations of the input images. We show that sharing these features with the segmentation head, i.e. decoder of the U-Net architecture improves the classification quality. Moreover, one can see in Tab.~\ref{tab:results} that this effect is enhanced by sharing the spatial feature maps from the upper levels of the U-Net's decoder. The proposed \textit{Multitask-Spatial-1} architecture (see Fig.~\ref{fig:model}) with shallow segmentation and classification heads directly sharing the same spatial feature map shows the top classification results. Especially, it most accurately distinguish between COVID-19 and other lung diseases (ROC-AUC COVID-19 vs Bac.~Pneum.~$= 0.87$, ROC-AUC COVID-19 vs Nodules~$=0.93$).

As seen in Tab.~\ref{tab:results} and Fig.~\ref{fig:contours} there is no significant difference in terms of segmentation and severity quantification qualities between the multitask models and the neural networks for single segmentation task.

Therefore, the single proposed \textit{Multitask-Spatial-1} model can be applied for both triage problems: identification of COVID-19 patients followed by their ranking according to the severity. In Fig.~\ref{fig:cls_rnk} we visualize these two steps of triage pipeline for the test dataset, described in Sec.~\ref{ssec:data:test}. One can see the several false positive alarms in cases with non-COVID-19 pathological findings. We discuss the possible ways to resolve them in Sec.~\ref{sec:discussion}. The overall pipeline for triage, including preprocessing, lungs segmentation, and multitask inference takes 8s and 20s using nVidia V100 and GTX 980 GPUs respectively.
}


%% file: content/6_discussion.tex
\section{Discussion}
\label{sec:discussion}
\begin{figure}[h]
    \centering
    \includegraphics[width=1\linewidth]{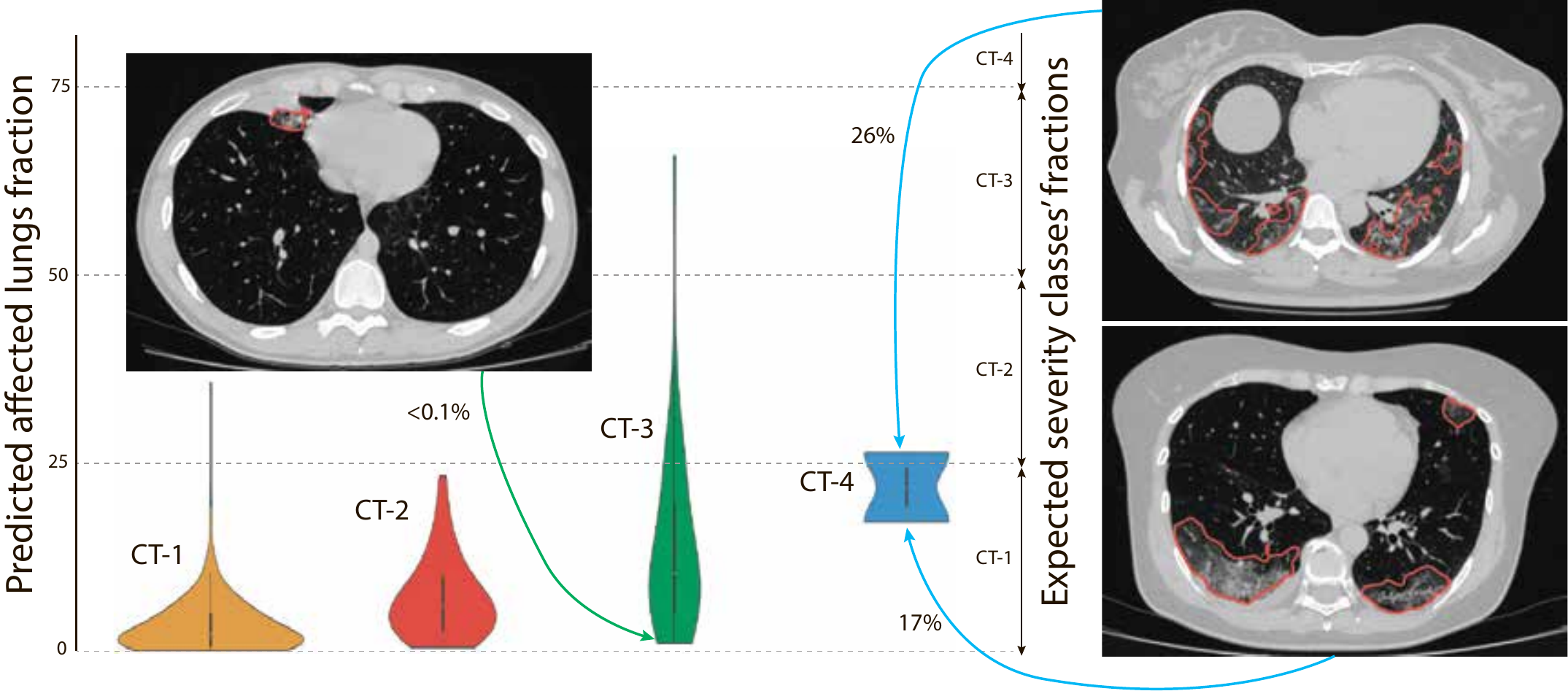}

    \caption{
        Predicted \textit{Severity} for weakly annotated cases from the Mosmed-1110 dataset. Note the inconsistency of visual subjective estimation.
        The lines denote the correspondence between some visually underestimated cases and their representative axial slices. 
        %
        \label{fig:subjective}
    }
\end{figure}

We have highlighted two important scores: COVID-19 \textit{Identification} and \textit{Severity} and discussed their priorities in different clinical scenarios. We have shown that these two scores aren't aligned well. Existing methods operate either with \textit{Identification} or \textit{Severity} and demonstrate deteriorated performance for the other task. We have presented a new method for joint estimation of COVID-19 \textit{Identification} and \textit{Severity} score and showed that the proposed multitask architecture achieves top quality metrics for both tasks simultaneously. Finally, we have released the code and used public data for training, so our results are fully reproducible.

\answer{However, as we show in Fig.~\ref{fig:cls_rnk} the proposed triage pipeline yields a relatively small number of false-positive alarms in patients with non-COVID-19 lungs diseases. But in practice, usage of an automated triage system always implies second reading. Therefore these false positives can be resolved by a radiologist. Moreover, it can be impossible to diagnose COVID-19 based only on CT image as PCR testing remains a gold-standard method. Thus, we conclude that the identification part may be used as a high sensitivity first reading tool to highlight patients with suspicious changes in their lungs. }

\answer{
The role of the  \textit{Severity Quantification} part is more straightforward. As we mentioned in Section \ref{sec:intro}, radiologists perform the severity classification into groups from CT0 (no COVID-19 related lesions) and CT1 (up to 25\% of lungs affected) to CT4 (more than 75\%) in a visual semi-quantitative fashion. We believe that such estimation may be highly subjective and may contain severe discrepancies. To validate this assumption, we additionally analyzed Mosmed-1110, which includes not only 50 segmentation masks but also 1110 multiclass labels CT0-CT4. Within our experiments, we binarized these labels and effectively removed information about COVID-19 severity. We examined mask predictions for the remaining 1050 cases, excluding healthy patients (CT0 group) and grouped the predictions by these weak labels, as shown in Fig. \ref{fig:subjective}. An expert radiologist validated analyzed the most extreme mismatches visualized in Fig. \ref{fig:subjective} and confirmed the correctness of our model's predictions.  As we see, the severity of many studies was highly underestimated during the visual semi-quantitative analysis. This result implies that deep-learning-based medical image analysis algorithms, including the proposed method, are great intelligent radiologists' assistants in a fast and reliable estimation of time-consuming biomarkers such as COVID-19 severity.}